\documentclass[prb,superscriptaddress,twocolumn,longbibliography]{revtex4-1}

\usepackage{amsmath,amssymb}
\usepackage{color}

\usepackage{tabularx}
\usepackage{ifpdf}

\ifpdf
  \usepackage[pdftex]{graphicx}
  \DeclareGraphicsExtensions{.pdf}
  \usepackage[hyperindex=true]{hyperref}
\else
  \usepackage{graphicx}
  \DeclareGraphicsExtensions{.ps,.eps}

  \usepackage[a4paper,dvipdfm,hyperindex=true]{hyperref}
\fi

\definecolor{linkcol}{rgb}{0.2,0.2,0.6}

\hypersetup{bookmarksopen=true,
pdfmenubar=true, 
pdfhighlight=/O, 
colorlinks=true, 
pdfpagemode=None, 
pdffitwindow=true, 
linkcolor=linkcol, 
citecolor=linkcol, 
urlcolor=linkcol 
}

\def\Qb{\ensuremath{|\mathbf{Q}|}}
\def\SCTO{Sr$_2$CuTeO$_6$}
\def\SCWO{Sr$_2$CuWO$_6$}
\def\SCTWOx{Sr$_2$CuTe$_{1-x}$W$_x$O$_6$}
\def\SCTWOf{Sr$_2$CuTe$_{0.5}$W$_{0.5}$O$_6$}

\newcolumntype{L}[1]{>{\raggedright\arraybackslash}p{#1}}
\newcolumntype{C}[1]{>{\centering\arraybackslash}p{#1}}
\newcolumntype{R}[1]{>{\raggedleft\arraybackslash}p{#1}}

\begin{document}

\title{Randomness and Frustration in a $S = 1/2$ Square-Lattice Heisenberg 
Antiferromagnet}

\author{Ellen Fogh}
\email{ellen.fogh@epfl.ch}
\affiliation{Laboratory for Quantum Magnetism, Institute of Physics, 
\'{E}cole Polytechnique F\'{e}d\'{e}rale de Lausanne (EPFL), CH-1015 Lausanne, 
Switzerland}
\author{Otto Mustonen}
\affiliation{Department of Materials Science and Engineering, University of 
Sheffield, Mappin Street, Sheffield, S1 3JD, United Kingdom}
\affiliation{School of Chemistry, University of Birmingham, Edgbaston, 
Birmingham B15 2TT, United Kingdom}
\author{Peter Babkevich}
\affiliation{Laboratory for Quantum Magnetism, Institute of Physics, 
\'{E}cole Polytechnique F\'{e}d\'{e}rale de Lausanne (EPFL), CH-1015 Lausanne, 
Switzerland}
\author{Vamshi M. Katukuri}
\affiliation{Max Planck Institute for Solid State Research, Heisenbergstr.~1, 
70569 Stuttgart, Germany}
\author{Helen C. Walker}
\affiliation{ISIS Neutron and Muon Source, Rutherford Appleton Laboratory, 
Chilton, Didcot, OX11 OQX, United Kingdom}
\author{Lucile Mangin-Thro}
\affiliation{Institute Laue Langevin, 71 Avenue des Martyrs, CS 20156, 38042 
Grenoble Cedex 9, France}
\author{Maarit Karppinen}
\affiliation{Department of Chemistry and Materials Science, Aalto University, 
FI-00076 Espoo, Finland}
\author{Simon Ward}
\affiliation{European Spallation Source ERIC, P.O. Box 176, SE-221 00, Lund, 
Sweden}
\author{Bruce Normand}
\affiliation{Paul Scherrer Institute, CH-5232 Villigen-PSI, Switzerland}
\affiliation{Laboratory for Quantum Magnetism, Institute of Physics, 
\'{E}cole Polytechnique F\'{e}d\'{e}rale de Lausanne (EPFL), CH-1015 Lausanne, 
Switzerland}
\author{Henrik M. R\o nnow}
\affiliation{Laboratory for Quantum Magnetism, Institute of Physics, 
\'{E}cole Polytechnique F\'{e}d\'{e}rale de Lausanne (EPFL), CH-1015 Lausanne, 
Switzerland}

\begin{abstract}

We explore the interplay between randomness and magnetic frustration in 
the series of $S = 1/2$ Heisenberg square-lattice compounds \SCTWOx. 
Substituting W for Te alters the magnetic interactions dramatically, 
from strongly nearest-neighbor to next-nearest-neighbor antiferromagnetic 
coupling. We perform neutron scattering measurements to probe the magnetic 
ground state and excitations over a range of $x$. We propose a bond-disorder 
model that reproduces ground states with only short-ranged spin correlations 
in the mixed compounds. The calculated neutron diffraction patterns and powder 
spectra agree well with the measured data and allow detailed predictions for 
future measurements. We conclude that quenched randomness plays the major role 
in defining the physics of \SCTWOx\ with frustration being less significant.

\end{abstract}

\date{\today}
\maketitle

\section{Introduction}
\begin{figure*}[t]
\includegraphics[width=\textwidth]{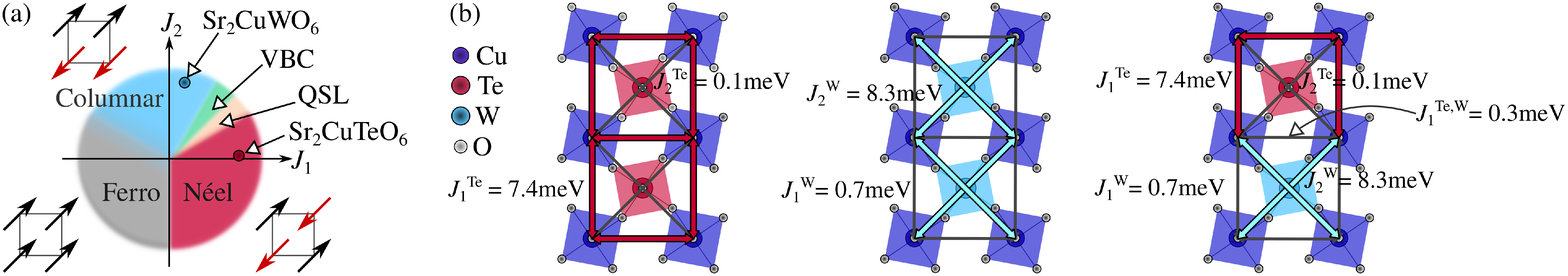}
\caption{{\bf The Heisenberg square lattice and \SCTWOx.} (a) Phase diagram 
of the $J_1$-$J_2$ square-lattice Heisenberg model, showing the Ferromagnetic, 
Columnar, and N\'eel AF states, as well as the frustrated parameter regimes 
of QSL and Valence-Bond Crystal (VBC) behavior. (b) Magnetic interactions in 
\SCTWOx, represented for the three cases where the counterions in each square 
are both Te, both W, or one of each. The $J_1$ and $J_2$ parameters are those 
obtained by quantum chemistry calculations \cite{vamshi_2020}; we note that 
the nearest-neighbor interaction is always small in the presence of W.}
\label{fig:bonds}
\end{figure*}

The Heisenberg $S = 1/2$ square lattice with competing antiferromagnetic 
(AF) nearest- and next-nearest-neighbor interactions, $J_1$ on the sides 
and $J_2$ on the diagonals of each square, presents a prototypical frustrated 
magnetic system \cite{chandra_1988,darradi_2008}. As Fig.~\ref{fig:bonds}(a) 
represents, when $J_1$ dominates the ground state is N\'eel AF order, whereas 
dominant $J_2$ gives a columnar AF state, and a quantum spin liquid (QSL) has 
been proposed \cite{morita_2015,gong_2014,wang_2018} within the non-ordered 
parameter regime ($0.4 \lesssim J_2/J_1 \lesssim 0.6$) between the two AF 
phases. Despite several decades of intense study, there remains no consensus 
over the exact nature of the QSL and the model continues to provide 
a focal point for QSL research.

Until recently, most such research in both experiment and theory was focused 
on homogenous systems, where all magnetic sites are equal. However, many real 
materials display intrinsic inhomogeneity, as a result of impurities or 
(counter-)ion substitution, that results in site or bond disorder. This 
is known as quenched randomness, and the loss of translational symmetry 
it entails makes the system challenging to study theoretically. However, 
quenched randomness in quantum magnets can lead to specific ground states 
with no long-ranged order, including the Bose glass \cite{rfwgf,rps,rakpr1,
rnwh,rmkg,Yu2012}, the Mott glass \cite{rogl,rakpr2,Yu2012,savary_2017}, the 
random-singlet state \cite{rbl,rdsf,liu_2018,uematsu_2018,ren_2020,baek_2020}, 
and the valence-bond glass \cite{tarzia_2008,singh_2010,watanabe_2018}. 
These phases of matter are closely related to certain types of QSL and thus 
raise the question of whether randomness in a frustrated system can produce 
qualitatively different types of quantum coherence, as opposed to only 
destroying such coherence. 

Here we investigate the magnetically disordered states 
found in the series of compounds \SCTWOx. At first sight this system 
seems well suited for exploring the phase diagram of the $J_1$-$J_2$ 
Heisenberg square lattice, because the two parent compounds, \SCTO\ 
and \SCWO, are respectively good $J_1$ and $J_2$ systems, displaying 
N\'eel and columnar AF order. However, our diffuse polarized neutron 
diffraction and inelastic neutron scattering (INS) measurements, combined 
with insight from mean-field and linear spin-wave calculations, show that 
the \SCTWOx\ family represents an altogether different but no less 
interesting problem. We demonstrate that the random-bond model arising 
from Te-W site disorder leads to a ground state of partially frozen moments 
in alternating patches of N\'eel and columnar correlations. The patch sizes 
depend on $x$, reaching a minimum of order 10 magnetic sites for $x = 0.4$. 
Our calculations reproduce well the experimentally observed ground and 
excited states, showing that disorder is more important than frustration 
in determining the physics of \SCTWOx.

\section{Materials}

The isostructural materials Sr$_2$Cu$B''$O$_6$ ($B''=\,$Te, W, Mo) are 
layered antiferromagnets in which the network of Cu$^{2+}$ ions is well 
described by the $J_1$-$J_2$ square-lattice model. Neutron scattering
measurements on large powder samples of the pure Te and W members of the 
family have shown that the ground state and spin dynamics of \SCTO\ are 
dominated by the $J_1$ term \cite{koga_2016,babkevich_2016}, whereas in 
\SCWO\ they are dominated by $J_2$ \cite{vasala_2014,walker_2016}. The fact 
that Te$^{6+}$ and W$^{6+}$ have almost identical ionic radii \cite{shannon_1976} 
gives every reason to expect that mixed compounds in the series between these 
two end members might realize ideal random solid solutions interpolating 
between the $J_1$ and $J_2$ limits. X-ray diffraction studies across the 
doping series \cite{mustonen_prb_2018} have established that the chemical 
structure of the mixed systems is indeed a true solid solution for all $x$, 
and detailed characterization of the magnetic response by muon spin-rotation 
($\mu$SR), specific-heat, magnetic susceptibility, and NMR measurements on 
\SCTWOf\ \cite{mustonen_ncomms_2018,mustonen_prb_2018,watanabe_2018} indicate 
no magnetic order above 19 mK over a wide range of doping, $0.1 \le x \le 
0.6$, which is clearly different from a two-phase system of the end members.

In the quest to understand the dramatic difference between \SCTO\ and \SCWO, 
\textit{ab initio} quantum chemistry calculations \cite{vamshi_2020} have 
demonstrated how the Cu-Cu superexchange paths change completely due to the 
orbital hybridization of O 2$p$ with Te$^{6+}$ (empty 5\textit{p}) or W$^{6+}$ 
(empty 5\textit{d}) \cite{shannon_1976}. The magnetic interaction parameters 
predicted by this analysis are shown in Fig.~\ref{fig:bonds}(b), and they 
afford the key insight that shapes both the physics of \SCTWOf\ and the 
applicability of our spin-wave methodology, namely that all the competing 
bonds are very weak and hence strong local frustration is avoided. Because 
the substitution of a nonmagnetic ion switches the dominant Cu-Cu interaction 
so cleanly, while leaving the crystal structure basically unaltered, the random 
Te-W distribution leads to a bond-disorder problem with bimodal distributions 
of $J_1$ and $J_2$. Among other things, the concept of controlling a uniform 
$J_2/J_1$ ratio to obtain a QSL by substitution is not valid. Nevertheless, 
one may still anticipate randomness-induced magnetic disorder, as suggested by 
magnetic measurements on samples spanning the doping range $0.1 \le x \le 0.6$,
\cite{mustonen_ncomms_2018,mustonen_prb_2018,watanabe_2018}, and recent 
studies have stressed the very rapid destruction of N\'eel order at small 
$x$ \cite{hong_2021,yoon_2021} (but not at high $x$ \cite{hong_2021}). These 
results have been interpreted theoretically in terms of a random-singlet 
state \cite{uematsu_2018,hong_2021} and of a valence-bond glass 
\cite{watanabe_2018}. However, INS measurements show dispersive excitations 
similar to spin waves \cite{vamshi_2020,hu_2021} and a partial freezing of 
random moments has been reported both at $x = 0.5$ below 1.7 K \cite{hu_2021} 
and at $x = 0.05$ below 0.5 K \cite{yoon_2021}. These somewhat contradictory 
findings leave the true nature of the magnetic ground state in \SCTWOx\ 
undetermined.

\begin{figure}[b]
\includegraphics[width = \columnwidth]{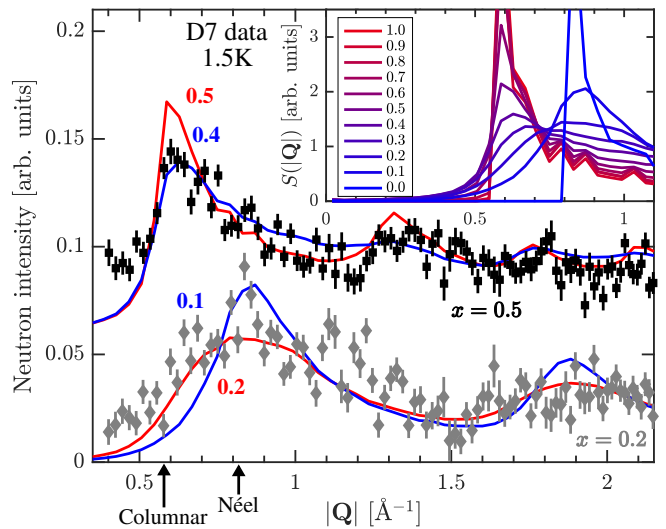}
\caption{{\bf Polarized neutron diffraction.} Intensities measured for powder 
samples of \SCTWOx\ with $x = 0.2$ (gray diamonds) and $x = 0.5$ (black 
squares). Data for $x = 0.5$ are translated upwards by 0.06 arb.~u.~for 
clarity. The inset shows intensities calculated using the random-bond model 
of Fig.~\ref{fig:bonds}(b); four of these curves are scaled and superimposed 
on the data in the main panel.}
\label{fig:diffraction}
\end{figure}

\section{Magnetic ground state}
The instantaneous spin structure factor, $S({\bf Q}) = \sum_j e^{i {\bf Q} 
\cdot {\bf r}_j} \left\langle {\bf S}_0 \cdot {\bf S}_j \right\rangle$, can be
probed by diffuse polarized neutron diffraction. Here ${\bf Q}$ is the momentum 
transfer and ${\bf S}_j$ the spin operator at lattice site ${\bf r}_j$. We 
collected diffraction patterns on the D7 diffractometer at the Institut Laue 
Langevin (ILL) using powder samples of 10-15 g of the $x = 0.2$ and $x = 0.5$ 
materials in Al cans at $T = 1.5$ K. The experimental wavelength of $4.8 \; 
\mathrm{\AA}$ corresponds to $3.55$ meV, and thus captures fluctuations in the 
lowest 20\% of the full band width \cite{babkevich_2016,walker_2016} by energy, 
which nevertheless constitute the vast majority of fluctuations by spectral 
weight (as we will show by INS). Thus our measurements observe the quasielastic 
response, or slowly fluctuating part of $S({\bf Q})$. The magnetic contribution 
to the scattering extracted by polarization analysis \cite{ehlers_2013} is 
shown in Fig.~\ref{fig:diffraction} and indicates a disordered state whose 
peak scattering intensity moves from $0.8 \; \mathrm{\AA^{-1}}$ for $x = 0.2$ 
to $0.6 \; \mathrm{\AA^{-1}}$ for $x = 0.5$, the former lying close to the 
magnetic Bragg peak $(0.5,0.5)$ of N\'eel order and the latter to $(0.5,0)$ 
of columnar order. 

\begin{figure}[p]
\includegraphics[width = \columnwidth]{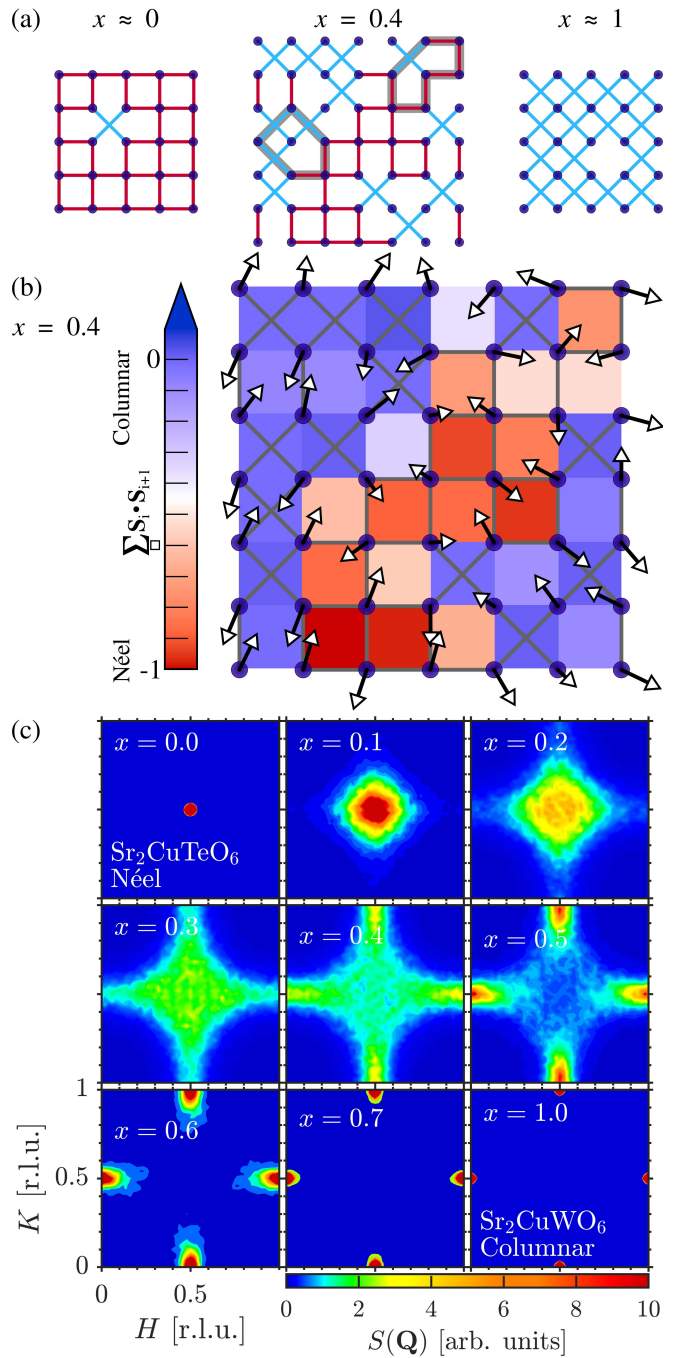}
\caption{{\bf Ground-state spin configurations.} (a) Examples of 
random-bond networks for Sr$_2$CuTe$_{1-x}$W$_x$O$_6$ with $x \approx 0$, 
$x = 0.4$, and $x \approx 1$; weak interactions are omitted for clarity. 
Thick gray lines at $x = 0.4$ indicate two representative geometrically 
frustrated paths, with five bonds being the shortest possible. (b) Spin 
configuration for $x = 0.4$, matching the bond network in panel (a). 
The color code quantifies the correlations around each square, 
$\sum_{\Box} {\bf S}_i \cdot {\bf S}_{i+1}$ ($i = \left\lbrace 1,2,3,4 
\right\rbrace$ labelling the four sites), which vary from N\'eel (red) to 
columnar (blue) character. Because the spins are of fixed size but rotate in 
three dimensions, shorter arrows indicate an out-of-plane component. (c) 
Calculated structure factor, $S(\mathbf{Q})$, showing the evolution from 
N\'eel to columnar order. Only for $x = 0$ and 1 are the peaks very sharp, 
as expected for long-range order.}
\label{fig:spinconfig}
\end{figure}

To interpret the diffraction data, the interaction parameters shown in 
Fig.~\ref{fig:bonds}(b) motivate a ground state for the \SCTWOx\ system 
whose essential behavior is captured by considering only the strongest bonds, 
i.e.~$J_1^{\mathrm{Te}}$ and $J_2^{\mathrm{W}}$. As Fig.~\ref{fig:spinconfig}(a) 
makes clear, the way that the introduction of W eliminates so many $J_1$ 
bonds leads to a somewhat dilute interaction network, and in particular we 
observe that direct $J_1$-$J_2$ frustration, in the form of $J_1$-$J_1$-$J_2$ 
triangles, is absent in this limit. Although the physics of the ground and 
excited states in the random \SCTWOx\ system can be understood directly 
from Fig.~\ref{fig:spinconfig}(a), we restore the weaker couplings in our 
quantitative modelling of both. For this we substitute the calculated 
interaction parameters \cite{vamshi_2020} shown in Fig.~\ref{fig:bonds}(b) 
by the rather similar values extracted from spin-wave fits to the INS spectra 
of \SCTO\ \cite{babkevich_2016} and \SCWO\ \cite{walker_2016}, which are 
$J_1^{\mathrm{Te}} = 7.6$ meV, $J_2^{\mathrm{Te}} = 0.6$ meV, $J_1^{\mathrm{W}} = 1.0$ 
meV, and $J_2^{\mathrm{W}} = 8.5$ meV, while the undetermined coupling 
$J_1^{\mathrm{Te,W}}$ is set to zero. 

The spin configuration corresponding to the random-bond network for $x = 0.4$ 
in Fig.~\ref{fig:spinconfig}(a) is illustrated in Fig.~\ref{fig:spinconfig}(b). 
We calculate these configurations by updating all sites in a random sequence, 
orienting spin $i$ to minimize its energy, $E_i = \sum_{j} J_{ij} \, {\bf S}_i \! 
\cdot \! {\bf S}_j = {\bf S}_i \! \cdot \! {\bf m}_i$, in the local classical 
mean field, ${\bf m}_i = \left\langle J_{ij} {\bf S}_j \right\rangle$, due to 
all neighboring spins, $j$. We repeat this procedure until the sum of 
differences in total energy from the previous 100 updates is below $10^{-9}$ 
meV for a system of 50$\times$50 sites. The spin magnitude is fixed to $\langle 
S \rangle = 1/2$ and the calculation is performed for 10 different random 
initial configurations at each value of $x$. We stress that all the spin 
configurations we obtain for $0 < x < 1$ are non-coplanar as a result of 
the randomness and the remaining frustration [marked in gray in 
Fig.~\ref{fig:spinconfig}(a)].

\begin{figure*}[t]
\centering
\includegraphics[width=1.0\textwidth]{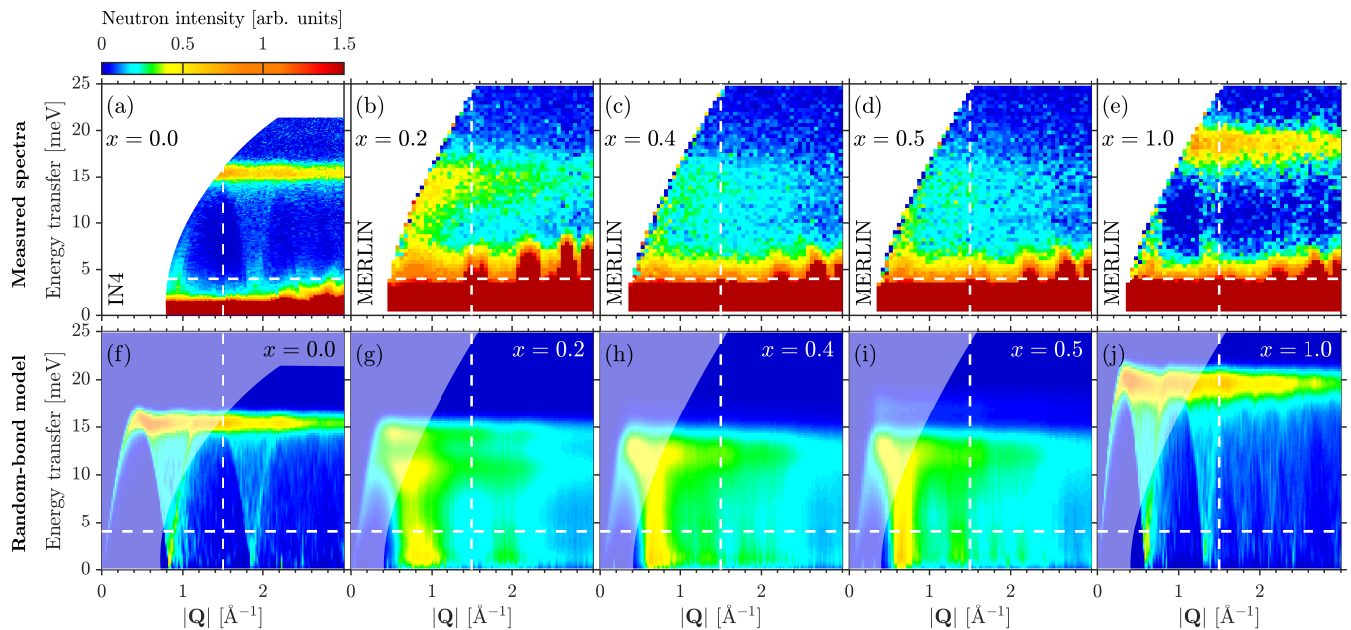}
\caption{{\bf Magnetic excitations in \SCTWOx.} (a-e) Powder INS spectra for 
samples with $x = 0$, 0.2, 0.4, 0.5, and 1, normalized to the nuclear Bragg 
peak for comparison between different $x$ values. (f-j) Spectra calculated 
using the random-bond model. The magnetic form factor for Cu$^{2+}$ was 
estimated following Ref.~\onlinecite{ITC} and an energy broadening ($dE = 1.2$ meV 
for $x = 0$ and 1.8 meV for $x > 0$) was convolved with each calculated 
spectrum to approximate the instrumental resolution. Regions of 
$|\mathbf{Q}|$ not covered in experiment are dimmed in the modelled 
spectra. The dashed lines mark the constant $|\mathbf{Q}|$ and energy 
values analyzed in Fig.~\ref{fig:SQvsQ}.}
\label{fig:spectra}
\end{figure*}

Averaged diffraction patterns derived from such spin structures are shown 
in Fig.~\ref{fig:spinconfig}(c). We find from the peaks at $(0.5,0)$ and 
equivalent positions that columnar ordering predominates for $x \gtrsim 0.5$. 
By contrast, the $(0.5,0.5)$ peak shows that long-ranged N\'eel order is 
destroyed even at $x \approx 0.1$ \cite{mustonen_prb_2018,hong_2021}. At 
intermediate $x$, only short-range magnetic correlations are present and the 
scattering is not isotropic, but forms instead a rounded, cross-like pattern 
in $S(\mathbf{Q})$. This indicates the presence of coexisting regions of very 
short-ranged N\'eel and columnar order, a real-space picture confirmed in 
Fig.~\ref{fig:spinconfig}(b). The sizes of these ``patches'' depend on $x$, 
and are of order 10 magnetic sites for $x = 0.4$. 

The powder averages of the diffraction patterns in Fig.~\ref{fig:spinconfig}(c) 
are shown in the inset of Fig.~\ref{fig:diffraction}. Because the spins in our 
mean-field calculations are static, the comparison with the slowly fluctuating 
part of $S({\bf Q})$, as probed by our D7 measurements, is fully justified. 
While it is clear that our model is entirely consistent with the observed 
diffuse diffraction signal, we cannot exclude other models on the basis of 
Fig.~\ref{fig:diffraction} alone. One prominent example is the response of 
sizeable magnetic domains of N\'eel and columnar order, which would give only 
broadened peaks at the Bragg positions in Fig.~\ref{fig:spinconfig}(c), but on 
powder-averaging would be difficult to distinguish from our measurements. 
However, such a superposition could not explain the complete lack of magnetic 
order observed by $\mu$SR and INS for $0.05 \le x \le 0.6$, and next we turn 
to our measurements of the spin dynamics to obtain further information.

\section{Spin dynamics}
Our INS experiments were performed at the time-of-flight 
spectrometer MERLIN \cite{merlin} at the ISIS Neutron and Muon Source, using 
10 g powder samples of the $x = 0.2$, 0.4, 0.5, and 1 materials in Al cans, 
with an incoming neutron energy of 45 meV and at a temperature of 5 K. In 
the spectra shown in Figs.~\ref{fig:spectra}(a-e), a thermally adjusted 
background factor (recorded above 100 K) was subtracted to 
remove the phonon contribution at larger \Qb.  The $x = 0$ data are those 
of Ref.~\onlinecite{babkevich_2016}. The pure compounds \SCTO\ and \SCWO\ show spin 
waves dispersing respectively from the zone centers of N\'eel and columnar 
order. Bands of strong scattering found around 16 meV for $x = 0$ and 18 meV 
for $x = 1$ correspond to van Hove singularities at the zone boundaries. 
Although the spectra of the mixed \SCTWOx\ compounds show strong broadening 
in momentum and energy transfer, both the low-energy dispersive features 
and the van Hove band remain present. 

\begin{figure}[t]
\centering
\includegraphics[width=\columnwidth]{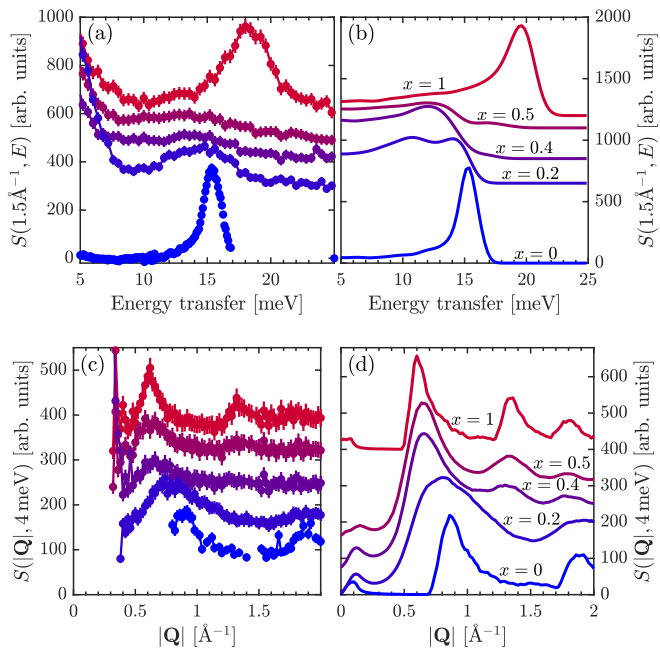}
\caption{{\bf Dynamical structure factor.} Measured (left) and calculated 
(right panels) $S(\Qb,E)$ at constant $\Qb = 1.5\;\mathrm{\AA^{-1}}$ (top) 
and constant $E = 4\;\mathrm{meV}$ (bottom panels). The respective 
integration widths are $\Delta \Qb = 0.2\;\mathrm{\AA^{-1}}$ and $\Delta 
E = 2$ meV. The curves are offset along the $y$-axis for clarity.}
\label{fig:SQvsQ}
\end{figure}

The success of the random-bond model in reproducing the ground states of the 
mixed systems (Fig.~\ref{fig:diffraction}) suggests its utility for analyzing 
their excitations. Despite having no long-range order, the mean-field states 
have frozen moments and thus we can compute powder-averaged INS 
spectra by using linear spin-wave theory, as implemented in the software 
package \textsc{spinw} \cite{spinW}. We define supercells of 10$\times$10 sites with 
random bond distributions of the type shown in Fig.~\ref{fig:spinconfig}(b) 
and periodic boundary conditions; for each value of $x$, we average the 
results from five different random configurations. For maximally quantitative 
modelling, we apply the series-expansion correction factor, $Z_c = 1.18$ 
\cite{singh_1989}, to our calculated energies. The resulting spectra, shown 
in Figs.~\ref{fig:spectra}(f-j), make clear that the random-bond model 
captures all the primary features of the measured spectra. Spin-wave-type 
modes, with substantial broadening in \Qb\ and energy, remain centered at 
$0.8\;\mathrm{\AA^{-1}}$ for $x = 0.2$, transitioning to $0.6\;\mathrm{\AA^{-1}}$ 
for $x = 0.5$, while the calculated band-width reduction is in quantitative 
agreement with the data. 

For further comparison, in Fig.~\ref{fig:SQvsQ}(a) we show a constant-\Qb\ 
cut through the spectrum at $\Qb = 1.5\;\mathrm{\AA^{-1}}$, where strong 
scattering is observed around 16 meV for \SCTO\ and 18 meV for \SCWO. For 
$x = 0.2$, 0.4, and 0.5, these van Hove peaks become a broad feature around 
12-15 meV that is reproduced accurately by our modelling, as shown in 
Fig.~\ref{fig:SQvsQ}(b). The increased scattering observed below 10 meV is 
the tail of the broadened elastic line, which we do not model. Similarly, 
a constant-energy cut at 4 meV, examined as a function of \Qb\ in 
Figs.~\ref{fig:SQvsQ}(c-d), captures the excitations dispersing 
upwards from the magnetic zone centers. For \SCTO\ and \SCWO, the first 
spin-wave branches emerge respectively from $\Qb = 0.8$ and $0.6\;
\mathrm{\AA^{-1}}$, while the excitations from the next Brillouin zone are 
found at $\Qb = 1.8$ and $1.3\;\mathrm{\AA^{-1}}$. As Te is replaced by W, 
these low-energy excitations change rapidly to resemble those of \SCWO, such 
that the spectra of the $x = 0.4$ and 0.5 compounds show the fingerprints 
mostly of the $x = 1$ system [both the $(0.5,0)$ and $(0.5,1.5)$ features 
strengthening from $x = 0.4$ to 0.5]. We stress that our modelling procedure 
has no free parameters, but clearly reproduces all the essential features of 
the \SCTWOx\ system at a semi-quantitative level. 

\section{Discussion and conclusions}
Thus we have shown that the \SCTWOx\ series realizes not highly frustrated 
magnetism but a random-bond model whose ground state at intermediate $x$ is 
a partially frozen disorder. This situation is a consequence of the strong 
suppression of $J_1$ bonds as soon as one neighboring Te site is substituted 
by W \cite{vamshi_2020}: Fig.~\ref{fig:spinconfig}(a) illustrates that, if 
one neglects the 10\% effect of the subdominant bonds, then no triangles 
are created and the shortest frustrated paths consist of five bonds, which 
furthermore are rather sparse within the bonding network. With such moderate 
geometric frustration, short-range order forms over patches considerably 
larger than the individual squares, as shown in Fig.~\ref{fig:spinconfig}(b). 
This partial frustration release explains why a semiclassical spin-wave and 
mean-field treatment captures the leading behavior of the maximally quantum 
mechanical $S = 1/2$ spin system reasonably well, even when more complex 
combinations of the different bonds ($J_1^{\rm Te}, J_2^{\rm Te}, J_1^{\rm W}, 
J_2^{\rm W}$) are used. Here we stress that quantum-fluctuation corrections to 
our analysis can act to reduce some local moments to zero, and future neutron 
diffraction studies could probe this site-dependent effect. Nevertheless, our 
experimental data and calculations suggest that the ground state is a 
percolating network of frozen, randomly oriented average moments, within 
which some sites and patches may be fully fluctuating. 

The possible existence of a weak frozen-moment phase is emerging as a key 
question in the understanding of bond randomness in $S = 1/2$ quantum magnets. 
Although early $\mu$SR measurements indicating no magnetic order down to sub-K 
temperatures for \SCTWOx\ with $0.1 \le x \le 0.6$, together with a $T$-linear 
specific heat, were suggested initially as evidence for a QSL state 
\cite{mustonen_prb_2018,mustonen_ncomms_2018}, further $\mu$SR investigations 
at very small $x$ are being interpreted \cite{hong_2021,yoon_2021} as evidence 
for the onset of a dominant random-singlet phase. The distinction between 
the random-singlet \cite{liu_2018,ren_2020,uematsu_2018} and valence-bond-glass 
phases \cite{watanabe_2018} is that all singlets in the latter state have 
finite gaps, whereas in the former they form a continuum of values to the 
gapless limit. However, INS data indicating a transition from liquid to weakly 
frozen behavior below 1.7 K for $x = 0.5$ \cite{hu_2021} and the $\mu$SR 
observation of a frozen component below 0.5 K at $x = 0.05$ \cite{yoon_2021} 
raise the question of whether a random frozen spin network can persist as an 
intermediate regime as long-range order is destroyed by bond randomness. 

Our results provide a qualitative basis on which to interpret these findings. 
Although both the diffuse diffraction pattern and the spin-wave-type 
excitations we observe are consistent with a random network, the finite 
momentum resolution in our experiment and the lack of quantum corrections 
in our model prevent an unambiguous conclusion. The fact that reports of 
a finite frozen moment are restricted to the low \cite{yoon_2021} and high 
\cite{hu_2021} ends of the substitution range for the suppression of 
long-range order suggests that $x$ may be an important factor in controlling 
the crossover to an entirely disordered (fluctuating, moment-free) ground 
state. The other key factor is likely to be the residual frustration, where 
our results suggest that \SCTWOx\ may lack the degree of frustration required 
to realize an unconventional all-singlet disordered state, but a system with 
stronger local frustration could indeed do so. On this note we stress that 
the \SCTWOx\ family remains an excellent framework for studying the interplay 
between randomness and frustration in $S = 1/2$ quantum magnets. The bond 
randomness of \SCTWOx\ also arises in systems including Ba$_2$Cu(Te,W)O$_6$ 
\cite{mustonen_2019,mustonen_2020} and Cr$_2$(Te,W)O$_6$ \cite{zhu_2014,
zhu_2015}, and we expect these compounds to allow a further investigation 
of exotic magnetic states at the nexus of disorder and frustration. 

In summary, we have presented neutron diffraction data and INS spectra for 
the $J_1$-$J_2$ square-lattice system \SCTWOx\ with $0 \leq x \leq 1$ and 
we have introduced a matching random-bond model based on \textit{ab initio} 
calculations. The model diffraction pattern in the magnetically disordered 
region ($0.1 < x < 0.5$) has a cross-like form that is fully consistent with 
our diffuse polarized neutron diffraction measurements. The ground state 
consists of small patches of predominantly nearest- or next-nearest-neighbor 
correlated spins dictated by the nonmagnetic dopant sites. Powder spectra 
obtained from the random-bond model reproduce the dispersive excitations and 
suppressed band maximum observed in our INS experiments. These findings show 
that it is the bond randomness in \SCTWOx, rather than the residual 
frustration, that drives the physics of the system.

\section*{Acknowledgments}
We thank E. Cussen, A. Sandvik, and O. Yazyev 
for helpful discussions. This work was funded by the Swiss National 
Science Foundation, including by its Sinergia networks MPBH (Grant 
Nos.~CRSII2\_141962/1 and CRSII2\_160765/1) and Nanoskyrmionics 
(Grant No.~171003), by the European Research Council through the project 
CONQUEST (Grant No.~259602) and the Synergy network HERO (Grant No.~810451), and by the 
Leverhulme Trust through Research Project Grant No.~RPG-2017-109 and Early 
Career Fellowship No.~ECF-2021-170. We thank the ILL and ISIS for allocating beamtime 
for this study. Data collected at both facilites are available as Refs.~\onlinecite{ILLdoi1,ILLdoi2,RB1520052,RB1620093,RB1890186}.


%

\end{document}